\documentclass[english,smallextended]{article}       

\usepackage[pdftex]{graphicx}
\usepackage{amssymb}
\usepackage{amsmath}
\usepackage{color}
\usepackage{authblk}

\usepackage{mciteplus}
\usepackage[hidelinks,bookmarksopen]{hyperref}
\begin{document}

\title{Coherence time analysis in semiconducting hybrid qubit under realistic experimental conditions}

\author[1]{E. Ferraro}
\author[1,2]{M. Fanciulli}
\author[1]{M. De Michielis}

\affil[1]{CNR-IMM, Unit of Agrate Brianza, Via C. Olivetti 2, 20864 Agrate Brianza (MB), Italy}
\affil[2]{Dipartimento di Scienza dei Materiali, University of Milano Bicocca, Via R. Cozzi 55, 20125 Milano, Italy}

\maketitle

\begin{abstract}
The unavoidable effect of the environmental noise due to nuclear spins and charge traps is included in the study of the hybrid qubit dynamics. Hybrid qubit dues its name to the advantageous combination of manipulation speed of a charge qubit with the longevity of a spin qubit. It consists of three electrons confined through external gate voltages in a double quantum dot and deserves special interest in quantum computation applications due to its advantages in terms of fabrication, control and only electrical manipulation. 
The fluctuations of the global magnatic field do not affect the dynamics of the hybrid qubit that is defined in a decoherence-free subspace. The main sources of decoherence come from local magnetic fluctuations and charge fluctuations. Our work is based on the hypothesis that gate voltages are fixed and keeped constant while qubit evolves in time. Coherence time of the hybrid qubit is extracted when model parameters take values achievable experimentally in the nuclear free isotope $^{28}$Si, natural Si and GaAs hosts. 
\end{abstract}

\section{Introduction}
Spin-based architectures in semiconducting hosts are intensely studied in view of potential applications in quantum computation and simulation \cite{Shulman-2012,Veldhorst-2014,Pla-2012,Maune-2012,Bluhm-2011,Tyryshkin-2012,RuiLi-2012,Coish-2005}. Quantum dot (QD)-based architectures \cite{Morton-2011,Veldhorst-2014,Kawakami-2014} and qubit realized through donor-atom nuclear or electron spins \cite{Klymenko-2015,Gamble-2015,Saraiva-2015,Fanciulli-2003,Fanciulli-2005} are deeply investigated over the years. In particular semiconductor based qubits assure long electron spin coherence times, easy manipulation, fast gate operations and potential for scaling \cite{Loss-1998,DiVincenzo-2000,Taylor-2005,Laird-2010}, in addition to the compatibility with the existing CMOS process. The hybrid qubit (HQ), in which the dominant mechanism of interaction is the exchange coupling between couple of three electrons confined in a double QD, has been proposed \cite{Shi-2012,Koh-2012} and broadly investigated from an experimental \cite{Kim-2014,Kim-2015,Thorgrimsson-2017} and theoretical \cite{Ferraro-2014,Ferraro-2015-qip,Ferraro-2017,DeMichielis-2015,Ferraro-2015-prb,Rotta-2016,Prati-npj,Wang-2017} point of view. 

The unavoidable environmental noise is the dominant cause of decoherence in the qubit dynamics and deeply affects the fidelity of quantum gates. The sources of disturbance come mainly from a magnetic source and an electrical one. HQ is only affected by local fluctuations (i.e. Overhauser fields) since it is a decoherence-free subspace qubit and then protected against global magnetic fluctuations. One source of disturbance is represented by the magnetic field due to the nuclear spins in the host material, moreover the fluctuations in the applied magnetic field that removes spin degeneracy of quantum states have to be considered. Magnetic noise strongly depends on the host material and thanks to the presence of stable isotopes with zero nuclear spins it is of minor entity in Si but becomes significant in GaAs compounds. The dynamical decoupling techniques and nuclear polarization are used to reduce the disturbance on the qubit and increase its coherence time. Charge noise instead originates from two sources: charge fluctuations on impurities that act as traps and fluctuations on the electrostatic gates adopted to confine the electrons, directly linked to the exchange couplings between the spins.    

The effects of magnetic and charge noises on the dynamics of the HQ is studied theoretically. The evolution in time of the HQ occurs keeping constant the external gate voltages. This choice makes easy the comparison between our theoretical results and the experimental ones due to the constant inputs to be applied by the experimentalist. The coherence times are extracted when different experimental conditions of interest are realized in the nuclear free isotope $^{28}$Si, natural Si and GaAs hosts in presence of both the environmental noises.

In Section 2 an overview on the HQ dynamics and the envelope-fitting procedure to extract coherence times in presence of noise are presented. In Section 3 an articulate study of the coherence times in the space of the HQ physical parameters in different semiconducting hosts is reported. Finally some concluding remarks in Section 4.

\section{Hybrid qubit noise model}
In the HQ three electrons are distributed during the operations between a double QD, with at least one electron in each. The qubit is effectively described combining the Hubbard-like model with a projector operator method \cite{Ferraro-2014} by the following Hamiltonian model in $\hbar$ units 
\begin{equation}\label{effmatrix} 
H=\frac{1}{2}E^z(\sigma_{1}^z+\sigma_{2}^z+\sigma_{3}^z)+\frac{1}{4}j'\boldsymbol{\sigma}_{1}\cdot\boldsymbol{\sigma}_{2}+\frac{1}{4}j_1\boldsymbol{\sigma}_{1}\cdot\boldsymbol{\sigma}_{3}+\frac{1}{4}j_2\boldsymbol{\sigma}_{2}\cdot\boldsymbol{\sigma}_{3}.
\end{equation}
In Eq.(\ref{effmatrix}), $\boldsymbol{\sigma}_i$ $(i=1,2,3)$ are the electron Pauli matrices and $E^z=g\mu_BB^z$ is the Zeeman energy associated to the magnetic field lying in the $\hat{z}$ direction. The constant $g$ is the electron g-factor and $\mu_B$ is the Bohr magneton. The coefficients $j'$, $j_1$ and $j_2$ are the effective exchange couplings among couple of electrons and include the dot tunneling, the dot bias and both on-site and off-site Coulomb interactions \cite{Ferraro-2014}. They are tunable acting on the tunneling couplings by external gates and on the inter-dot bias voltage. The Coulomb energy as well as the intra-dot bias voltage, directly linked to $j'$, are fixed by the geometry of the qubit.

HQ is encoded within the two-dimensional subspace with total angular momentum state $S=1/2$ and projection along $\hat{z}$, $S_z=-1/2$ \cite{Shi-2012}. The logical basis 
\begin{equation}\label{01}
|0\rangle\equiv|S\rangle|\!\downarrow\rangle, \qquad |1\rangle\equiv\sqrt{\frac{1}{3}}|T_0\rangle|\!\downarrow\rangle-\sqrt{\frac{2}{3}}|T_-\rangle|\!\uparrow\rangle
\end{equation}
is composed by singlet ($|S\rangle$) and triplet ($|T_0\rangle$ and $|T_-\rangle$) states of a pair of electrons in combination with the single angular momentum states ($|\!\uparrow\rangle$ and $|\!\downarrow\rangle$) of the third spin, through Clebsh-Gordan coefficients. The dynamics is studied in the $2\times 2$ logical basis and the evolution operator $U(t)=e^{-iHt}$, where $H$ is given in Eq.(\ref{effmatrix}), is obtained analytically \cite{Ferraro-2018}.

Once that the initial condition $|\psi(0)\rangle$ is fixed and the system is free to evolve in time, the return probability $P_{|\varphi\rangle}(t)$ of finding the qubit in a given logical state $|\varphi\rangle$ at an arbitrary time instant $t$ is calculated exploiting the compact form of the evolution operator \cite{Ferraro-2018}. Employing the quasi-static model, both the magnetic disturbance and the electrical one are evaluated looking at the disorder-averaged return probability \cite{DasSarma-2016,DasSarma-2017,Wu-2017} 
\begin{equation}\label{average}
[P_{|\varphi\rangle}(t)]_{\alpha}=\int_0^{+\infty}\int_0^{+\infty}\int_{-\infty}^{+\infty}dj_1dj_2d(\delta E)f_{\delta E}(\delta E)f_{j_1}(j_1)f_{j_2}(j_2)P_{|\varphi\rangle}(t).
\end{equation}
The disturbance dues to local magnetic field fluctuations acting between the two QDs obeys to a Gaussian distribution $f_{\delta E}(\delta E)$ with zero mean and standard deviation $\sqrt{2}\sigma_E$ 
\begin{equation}
f_{\delta E}(\delta E)=\frac{1}{2\sigma_E\sqrt{\pi}}e^{-\frac{(\delta E)^2}{4\sigma_E^2}}.
\end{equation}
Analogously the exchange couplings $j_1$ and $j_2$ follow a Gaussian distribution $f_{j_i}(j_i)$, $(i=1,2)$, restricted to non-negative values with mean $j_{0i}$ and standard deviation $\sigma_{j_i}$
\begin{equation}
f_{j_i}(j_i)=\frac{1}{\sigma_{j_i}\sqrt{2\pi}}\frac{2}{1+erf(\frac{j_{0i}}{\sigma_{j_i}\sqrt{2}})}e^{-\frac{(j_i-j_{0i})^2}{2\sigma_{j_i}^2}}.
\end{equation}
The intra-dot exchange coupling $j'$ is mostly set by the geometry of the system and only less effectively tuned from external gates \cite{Ferraro-2014}. Due to this difference, in the following we are assuming that $j'$ is constant.

In the following the initial condition corresponding to the pure zero logical state $|\psi(0)\rangle=|0\rangle$ is studied. We are interested in the probability $P_{|0\rangle}(t)$ of finding the qubit in the $|0\rangle$ logical state that is an oscillatory function whose analytical expression at an arbitrary time instant $t$ is given by
\begin{equation}\label{P1}
P_{|0\rangle}(t)=1-\frac{4C^2}{(A-B)^2+4C^2}\sin^2(\beta t),
\end{equation}
where 
\begin{align}\label{coeff1}
&A=\frac{E^z}{2}+\frac{3}{4}j',\nonumber\\
&B=\frac{E^z}{2}-\frac{1}{4}j'+\frac{1}{2}(j_1+j_2),\nonumber\\
&C=\frac{\sqrt{3}}{4}(j_1-j_2)
\end{align}
and
\begin{equation}\label{coeff2}
\beta=\frac{\sqrt{(A-B)^2+4C^2}}{2}.
\end{equation}

The multiple integrals $[P_{|0\rangle}(t)]_{\alpha}$, in correspondence to several values of $\sigma_E$ and $\sigma_{j_1}=\sigma_{j_2}\equiv\sigma_j$, obtained inserting Eq.(\ref{P1}) into Eq.(\ref{average}) are evaluated numerically. From $[P_{|0\rangle}(t)]_{\alpha}$ coherence times $T_2^{\ast}$ are extracted for several experimental conditions of interest. 

Generally speaking $T_2^{\ast}$ takes into account the number of coherent oscillations shown by $[P_{|\varphi\rangle}(t)]_{\alpha}$ before it decays and it is evaluated through an envelope-fitting procedure in which we look for a curve of the form
\begin{equation}\label{fit_eq}
\mathcal{P}(t)=\mathcal{P}(0)+\left(1-\mathcal{P}(0)\right)e^{-(t/T_2^{\ast})^{\alpha}},
\end{equation}
that closely approximates the envelope function of $[P_{|0\rangle}(t)]_{\alpha}$. 

\section{Coherence time analysis}
In this Section a comprehensive study on the HQ coherence times is presented. The results are obtained exploring a suitable range of qubit physical parameters. 

The exchange couplings varies with the detuning $\varepsilon$ between the two quantum dots and can be generally described by the exponential function $j_i(\varepsilon)=j_{0i}e^{-\frac{\varepsilon}{\varepsilon_0}}\, (i=1,2)$ \cite{XianWu-2014}, where $\varepsilon_0$ is a detuning operating point taken as reference. Developing in power series and truncating, it follows that $\delta j_i\approx j_{0i}\frac{\delta\varepsilon}{\varepsilon_0}$ and consequently $\sigma_{j_i}\approx j_{0i}\frac{\sigma_\varepsilon}{\varepsilon_0}$. We work in the hypothesis that $j_{01}=0.5 j_0$, $j_{02}=1.5 j_0$ and $j'=0.5 j_0$. The parameter $j_0$ contains all the essential information in order to describe the qubit under investigation in terms of the qubit geometrical parameters and of the physical hosts materials where the qubit lies. Spin density functional theory allows us to simulate reasonable physical values for the parameter $j_0$ \cite{DeMichielis-2015}.

We can take advantage of the fact that $\sigma_E$ essentially remains constant as one changes the average exchange coupling $j_0$, while $\sigma_{j_i}$ is roughly linear in $j_0$. This assumption is less restrictive with respect to that one used in our previous work [33], where standard deviation $\sigma_j$ was considered as a simple independent variable with no explicit link to $j_0$. In the following the results are presented when $\sigma_{j_1}=\sigma_{j_2}\equiv\sigma_j$.

\begin{figure}[htbp]
	\begin{center}
\includegraphics[width=0.6\textwidth]{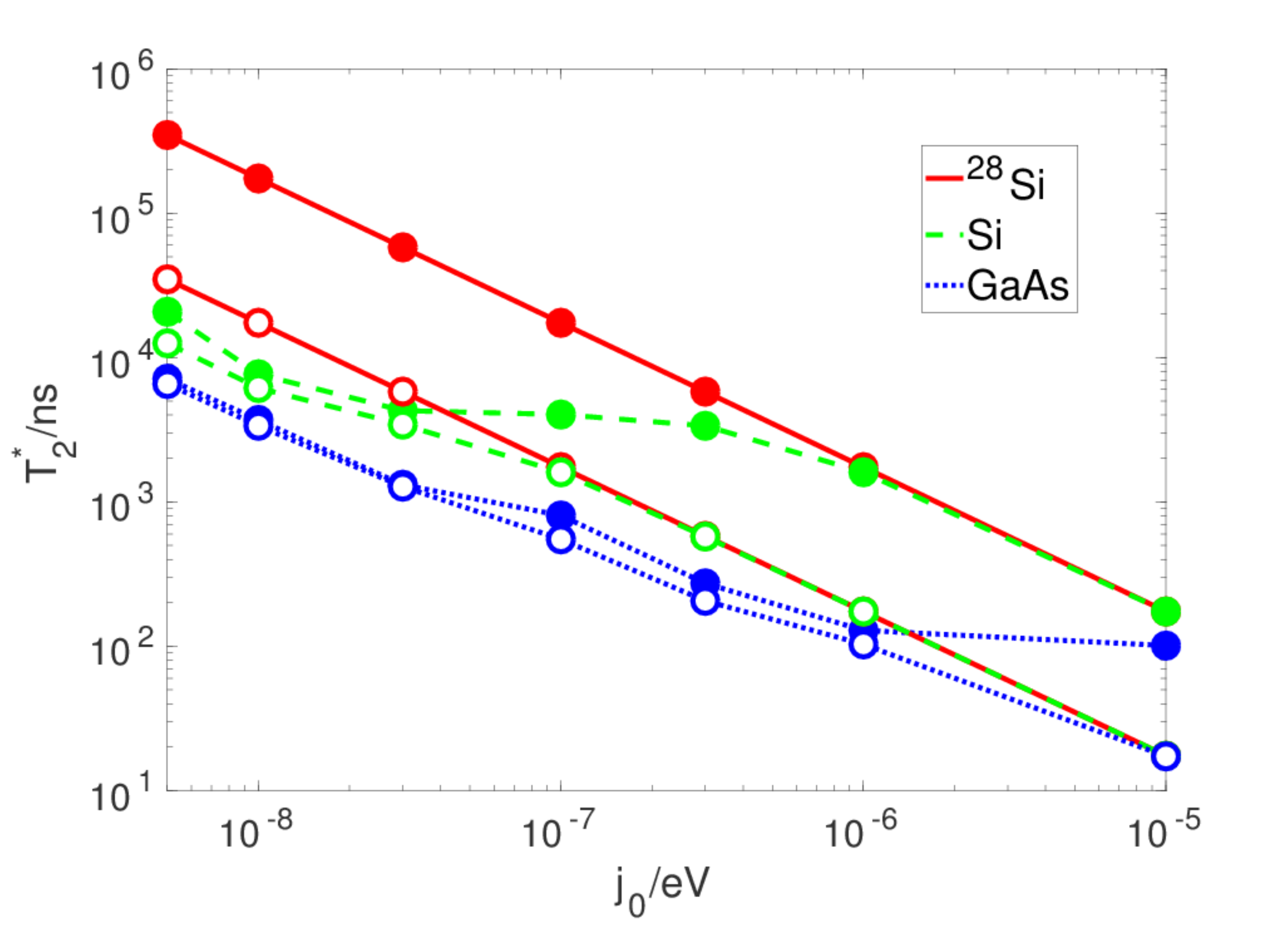}
	\end{center}
	\caption{$T_2^{\ast}$ as a function of $j_0$ when $\sigma_E=0$ ($^{28}$Si, red solid line), 3 neV (Si, green dashed line) and 100 neV (GaAs, blue dotted line). Two values of $\sigma_{\varepsilon}/\varepsilon_0=0.003$ (filled marks) and $\sigma_{\varepsilon}/\varepsilon_0=0.03$ (open marks) are studied.}\label{comparison1} 
\end{figure}

We compare in Fig. \ref{comparison1} the extracted coherence times from $[P_{|0\rangle}(t)]_{\alpha}$ through Eq.(\ref{fit_eq}) as a function of $j_0$. The parameter $j_0$ spans the range $[5\times 10^{-9}, 1\times 10^{-5}]$ eV in correspondence to two different values of $\sigma_{\varepsilon}/\varepsilon_0=0.003$ (filled marks) and $0.03$ (open marks). This choice is motivated by realistic experimental conditions and compatible with $\sigma_\varepsilon= 4.39$ $\mu$eV as reported in Ref. \cite{Thorgrimsson-2017}. Three different HQ semiconducting hosts are considered: $^{28}$Si ($\sigma_E=0$), Si ($\sigma_E=$ 3 neV)  and GaAs ($\sigma_E=$ 100 neV). 

All the curves in correspondence to a major source of charge disturbance ($\sigma_{\varepsilon}/\varepsilon_0=0.03$) lies under the curves with a minor one ($\sigma_{\varepsilon}/\varepsilon_0=0.003$) since, as expected, coherence times decrease. When large values for $j_0$ ($j_0$$\geq$ 10 $\mu$eV) are considered, charge noise dominates and the different nature of the host materials do not affect the HQ coherence times. Decreasing $j_0$, instead, $T_2^{\ast}$ increases and it is possible to appreciate how the Si and GaAs cases detach from the $^{28}$Si case. A long coherence time $T_2^{\ast}$ is a mandatory feature when the exploitation of the qubit memory aspects is desired. Moreover, focusing on Si, a \emph{plateau} region in which $T_2^{\ast}$ is constant in correspondence to $\sigma_{\varepsilon}/\varepsilon_0=0.003$ and $3\times 10^{-8}\leq j_0\leq 3\times 10^{-7}$ eV is present. In this range, a larger value of $j_0$ has to be preferred in order to assure a greater number of gate operations at an almost constant $T_2^{\ast}$. Note that the GaAs case shows a similar \emph{plateau} at $j_0$ $\geq$ 1 $\mu$eV. Same considerations can be done for the Si and GaAs cases with $\sigma_{\varepsilon}/\varepsilon_0=0.03$ where changes in the slope of the curves are visible but less evident. At lower $j_0$ the magnetic noise dominates and, in fact, the Si results lean roughly on the same curve independently on the level of charge noise. The same holds for GaAs case.

Our results suggest that a reduction in $j_0$ leads to a $T_2^{\ast}$ increase, when constant electrical potentials are applied ('always on' operation). A fair comparison between our foreseen $T_2^{\ast}$ and other predicted coherence times reported in literature is not straightforward due to the different type of control methods and sequences considered. In Ref. \cite{Thorgrimsson-2017}, $T_2^{\ast}$ obtained with various sequences in Si hybrid qubit is presented, and it is shown that higher coherence times can be achieved if the detuning amplitude $\varepsilon$ of the pulsed signal is increased. These results are not in contrast with our outcomes, even if our search was limited to a change only in $j_0$, because for a $\varepsilon$ increase a reduction in $j_0$ is obtained. A different implementation of the hybrid qubit in GaAs is presented in Ref. \cite{Wang-2017}, where three quantum dots are used instead of two. The authors performed initialization, control and readout of the qubit and presented a model to qualitatively understand the observed behaviours. As in our case, a decrease in the coherence time is directly linked to an increase in the exchange coupling between the electrons confined in the two dots where the hybrid qubit is defined. 

In order to compare our findings with the current experimental values, the experimental $T_2^{\ast}$ collected from the literature are reported for experiments with DC pulsed signals. For the Si case in Ref. \cite{Kim-2014} the authors estimate a lower bound for the coherence time $T_2^{\ast}$ for pulsed Larmor and Ramsey sequences of 2 and 10 ns, respectively. For the GaAs case in Ref. \cite{Wang-2017}, observed coherent oscillations, interpreted as Larmor oscillations, give a range of values for $T_2^{\ast}$ of 1-6 ns. Our $T_2^{\ast}$ results are beyond the experimental ones in the range of $j_0$ considered, with a reduced difference for high $j_0$, suggesting that the experimental working points are probably different with respect to the choice done in our calculations. More details on the experimental aspects of the system, such as charge noise levels, working point and exchange energies of the qubit, are required to better foresee the $T_2^{\ast}$ achievable. Note that a complete search of sweet spots for the hybrid qubit in the whole ranges of $j_1$, $j_2$ and $j'$ parameters will be a subject of a future work.

\begin{figure}[htbp]
	\begin{center}
\includegraphics[width=0.6\textwidth]{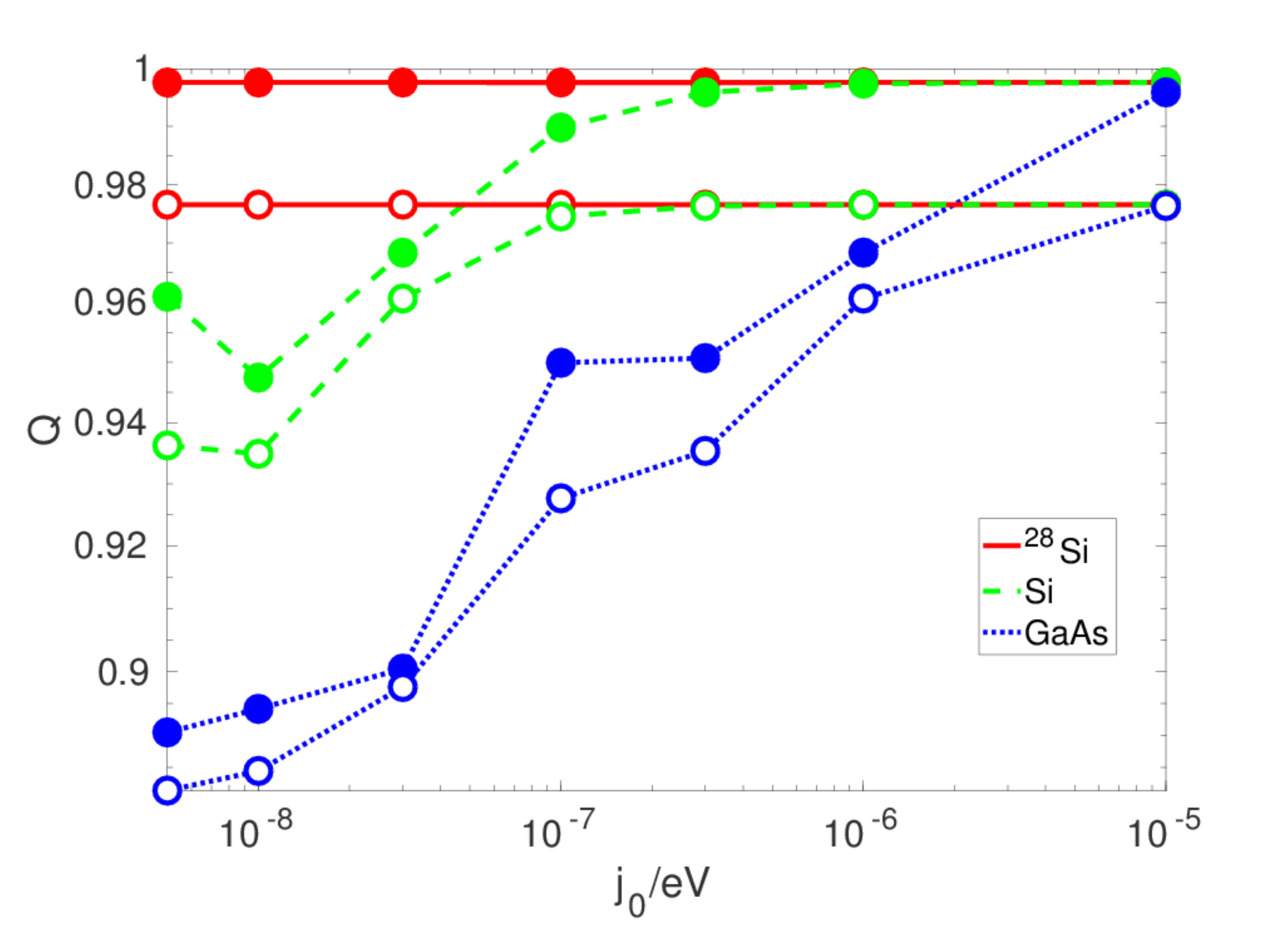}
	\end{center}
	\caption{The quality factor $Q$ as a function of $j_0$. The parameters are chosen as in Fig. \ref{comparison1}.} \label{comparison4} 
\end{figure}

Other interesting information can be extracted from the quality factor $Q$ \cite{DasSarma-2017}. It is a function of the coherence time $T_2^{\ast}$ and $j_0$ 
\begin{equation}
Q=\exp\left(-\frac{h}{j_0 T_2^{\ast}}\right)
\end{equation}
and represents the exponential decay factor for the return probability over $h/j_0$. Looking at the evolution of $Q$ with respect to $j_0$ becomes extremely important when the interest points on evaluating the best experimental condition to maximize the number of HQ state oscillations (gate operations).
Fig. \ref{comparison4} reports the behaviour of the quality factor $Q$ when $j_0$ is varied.
Differently from the $^{28}$Si case that is constant for every value of $j_0$ and is much closer to 1 when $\sigma_{\varepsilon}/\varepsilon_0=0.003$ with respect the case $\sigma_{\varepsilon}/\varepsilon_0=0.03$, $Q$ increases as a function of $j_0$ in Si and GaAs cases. At high $j_0$ the quality factors go toward the values of the corresponding $^{28}$Si case. Si case reaches the corresponding maxima $Q$ at lower $j_0$ (around 0.3 $\mu$eV) than the ones for the GaAs case (10 $\mu$eV). 

\section{Conclusions}
Hybrid qubit coherence times are extracted when both magnetic and electrical environmental noises are considered. The HQ time evolution is analytically derived keeping constant the control parameters. This means that from an experimental point of view, in order to make the comparison with our results, constant inputs have to be applied to the qubit. The coherence time is evaluated through the number of coherent oscillations shown by the disordered return probability before it decays and extracted adopting an envelope-fitting procedure. The in depth analysis of the behaviour of the coherence time with respect to $j_0$, the parameter that fully describes the qubit in terms of geometrical parameters and of the physical hosts materials, is of great interest in view of applications of the HQ in quantum computation. When $j_0\geq 10$ $\mu$eV, charge noise dominates and the different nature of the host materials do not affect the HQ coherence times. Decreasing $j_0$ the magnetic noise prevails and this is evident for Si below $j_0=30$ neV and for GaAs below $j_0=1 \mu$eV.

To complete the study, the behaviour of the quality factor Q with respect to $j_0$ is presented. It provides useful information when the interest points in identifying the best experimental condition to maximize the number of HQ state oscillations, directly related to the gate operations. $Q$ increases as a function of $j_0$ in Si and GaAs cases and at high $j_0$ it goes toward the values of the corresponding $^{28}$Si case that instead is constant for every value of $j_0$.

\section*{Acknowledgments}
This work has been funded from the European Union's Horizon 2020 research and innovation programme under grant agreement No 688539.

\bibliographystyle{angewandte}
\bibliography{Ref}

\ifx\mcitethebibliography\mciteundefinedmacro
\PackageError{angewandte.bst}{mciteplus.sty has not been loaded}
{This bibstyle requires the use of the mciteplus package.}\fi
\begin{mcitethebibliography}{10}
\newcommand{\enquote}[1]{``#1''}
\expandafter\ifx\csname urlstyle\endcsname\relax
  \providecommand{\doi}[1]{DOI \discretionary{}{}{}#1}\else
  \providecommand{\doi}{DOI \discretionary{}{}{}\begingroup
  \urlstyle{rm}\Url}\fi

\bibitem{Shulman-2012}
M.~D. Shulman, O.~E. Dial, S.~P. Harvey, H.~Bluhm, V.~Umansky, A.~Yacoby,
  \emph{Science} \textbf{2012}, \emph{336}, 202\relax
\mciteBstWouldAddEndPuncttrue
\mciteSetBstMidEndSepPunct{\mcitedefaultmidpunct}
{\mcitedefaultendpunct}{\mcitedefaultseppunct}\relax
\EndOfBibitem
\bibitem{Veldhorst-2014}
M.~Veldhorst, J.~C.~C. Hwang, C.~H. Yang, A.~W. Leenstra, B.~de~Ronde, J.~P.
  Dehollain, J.~T. Muhonen, F.~E. Hudson, K.~M. Itoh, A.~Morello, A.~S. Dzurak,
  \emph{Nature Nanotechnology} \textbf{2014}, \emph{9}, 981--985\relax
\mciteBstWouldAddEndPuncttrue
\mciteSetBstMidEndSepPunct{\mcitedefaultmidpunct}
{\mcitedefaultendpunct}{\mcitedefaultseppunct}\relax
\EndOfBibitem
\bibitem{Pla-2012}
J.~J. Pla, K.~Y. Tan, J.~P. Dehollain, W.~H. Lim, J.~J.~L. Morton, D.~N.
  Jamieson, A.~S. Dzurak, A.~Morello, \emph{Nature} \textbf{2012}, \emph{489},
  541\relax
\mciteBstWouldAddEndPuncttrue
\mciteSetBstMidEndSepPunct{\mcitedefaultmidpunct}
{\mcitedefaultendpunct}{\mcitedefaultseppunct}\relax
\EndOfBibitem
\bibitem{Maune-2012}
B.~M. Maune, M.~G. Borselli, B.~Huang, T.~D. Ladd, P.~W. Deelman, K.~S.
  Holabird, A.~A. Kiselev, I.~Alvarado-Rodriguez, R.~S. Ross, A.~E. Schmitz,
  M.~Sokolich, C.~A. Watson, M.~F. Gyure, A.~T. Hunter, \emph{Nature}
  \textbf{2012}, \emph{481}, 344--347\relax
\mciteBstWouldAddEndPuncttrue
\mciteSetBstMidEndSepPunct{\mcitedefaultmidpunct}
{\mcitedefaultendpunct}{\mcitedefaultseppunct}\relax
\EndOfBibitem
\bibitem{Bluhm-2011}
H.~Bluhm, S.~Foletti, I.~Neder, M.~Rudner, D.~Mahalu, V.~Umansky, A.~Yacoby,
  \emph{Nature Physics} \textbf{2011}, \emph{7}, 109\relax
\mciteBstWouldAddEndPuncttrue
\mciteSetBstMidEndSepPunct{\mcitedefaultmidpunct}
{\mcitedefaultendpunct}{\mcitedefaultseppunct}\relax
\EndOfBibitem
\bibitem{Tyryshkin-2012}
A.~M. Tyryshkin, S.~Tojo, J.~J.~L. Morton, H.~Riemann, N.~V. Abrosimov,
  P.~Becker, H.-J. Pohl, T.~Schenkel, M.~L.~W. Thewalt, K.~M. Itoh, S.~A. Lyon,
  \emph{Nature Materials} \textbf{2012}, \emph{11}, 143\relax
\mciteBstWouldAddEndPuncttrue
\mciteSetBstMidEndSepPunct{\mcitedefaultmidpunct}
{\mcitedefaultendpunct}{\mcitedefaultseppunct}\relax
\EndOfBibitem
\bibitem{RuiLi-2012}
R.~Li, X.~Hu, J.~Q. You, \emph{Physical Review B} \textbf{2012}, \emph{86},
  205306\relax
\mciteBstWouldAddEndPuncttrue
\mciteSetBstMidEndSepPunct{\mcitedefaultmidpunct}
{\mcitedefaultendpunct}{\mcitedefaultseppunct}\relax
\EndOfBibitem
\bibitem{Coish-2005}
W.~A. Coish, D.~Loss, \emph{Physical Review B} \textbf{2005}, \emph{72},
  125337\relax
\mciteBstWouldAddEndPuncttrue
\mciteSetBstMidEndSepPunct{\mcitedefaultmidpunct}
{\mcitedefaultendpunct}{\mcitedefaultseppunct}\relax
\EndOfBibitem
\bibitem{Morton-2011}
J.~J.~L. Morton, D.~R. McCamey, M.~A. Eriksson, S.~A. Lyon, \emph{Nature}
  \textbf{2011}, \emph{479}, 345--353\relax
\mciteBstWouldAddEndPuncttrue
\mciteSetBstMidEndSepPunct{\mcitedefaultmidpunct}
{\mcitedefaultendpunct}{\mcitedefaultseppunct}\relax
\EndOfBibitem
\bibitem{Kawakami-2014}
E.~Kawakami, P.~Scarlino, D.~R. Ward, F.~R. Braakman, D.~E. Savage, M.~G.
  Lagally, M.~Friesen, S.~N. Coppersmith, M.~A. Eriksson, L.~M.~K. Vandersypen,
  \emph{Nature Nanotechnology} \textbf{2014}, \emph{9}, 666--670\relax
\mciteBstWouldAddEndPuncttrue
\mciteSetBstMidEndSepPunct{\mcitedefaultmidpunct}
{\mcitedefaultendpunct}{\mcitedefaultseppunct}\relax
\EndOfBibitem
\bibitem{Klymenko-2015}
M.~V. Klymenko, S.~Rogge, F.~Remacle, \emph{Physical Review B} \textbf{2015},
  \emph{92}, 195302\relax
\mciteBstWouldAddEndPuncttrue
\mciteSetBstMidEndSepPunct{\mcitedefaultmidpunct}
{\mcitedefaultendpunct}{\mcitedefaultseppunct}\relax
\EndOfBibitem
\bibitem{Gamble-2015}
J.~K. Gamble, N.~T. Jacobson, E.~Nielsen, A.~D. Baczewski, J.~E. Moussa,
  I.~Montano, R.~P. Muller, \emph{Physical Review B} \textbf{2015}, \emph{91},
  235318\relax
\mciteBstWouldAddEndPuncttrue
\mciteSetBstMidEndSepPunct{\mcitedefaultmidpunct}
{\mcitedefaultendpunct}{\mcitedefaultseppunct}\relax
\EndOfBibitem
\bibitem{Saraiva-2015}
A.~L. Saraiva, A.~Baena, M.~J. Calder\'{o}n, B.~Koiller, \emph{J. Phys.:
  Condens. Matter} \textbf{2015}, \emph{27}, 154208\relax
\mciteBstWouldAddEndPuncttrue
\mciteSetBstMidEndSepPunct{\mcitedefaultmidpunct}
{\mcitedefaultendpunct}{\mcitedefaultseppunct}\relax
\EndOfBibitem
\bibitem{Fanciulli-2003}
M.~Fanciulli, P.~H\"ofer, A.~Ponti, \emph{Physica B} \textbf{2003},
  \emph{340-342}, 895--902\relax
\mciteBstWouldAddEndPuncttrue
\mciteSetBstMidEndSepPunct{\mcitedefaultmidpunct}
{\mcitedefaultendpunct}{\mcitedefaultseppunct}\relax
\EndOfBibitem
\bibitem{Fanciulli-2005}
A.~Ferretti, M.~Fanciulli, A.~Ponti, A.~Schweiger, \emph{Physical Review B}
  \textbf{2005}, \emph{72}, 235201\relax
\mciteBstWouldAddEndPuncttrue
\mciteSetBstMidEndSepPunct{\mcitedefaultmidpunct}
{\mcitedefaultendpunct}{\mcitedefaultseppunct}\relax
\EndOfBibitem
\bibitem{Loss-1998}
D.~Loss, D.~P. DiVincenzo, \emph{Physical Review A} \textbf{1998}, \emph{57},
  120\relax
\mciteBstWouldAddEndPuncttrue
\mciteSetBstMidEndSepPunct{\mcitedefaultmidpunct}
{\mcitedefaultendpunct}{\mcitedefaultseppunct}\relax
\EndOfBibitem
\bibitem{DiVincenzo-2000}
D.~P. DiVincenzo, D.~Bacon, J.~Kempe, G.~Burkard, , K.~B. Whaley, \emph{Nature
  (London)} \textbf{2000}, \emph{408}, 339\relax
\mciteBstWouldAddEndPuncttrue
\mciteSetBstMidEndSepPunct{\mcitedefaultmidpunct}
{\mcitedefaultendpunct}{\mcitedefaultseppunct}\relax
\EndOfBibitem
\bibitem{Taylor-2005}
J.~M. Taylor, H.-A. Engel, W.~D\"ur, A.~Yacoby, C.~M. Marcus, P.~Zoller, M.~D.
  Lukin, \emph{Nature Physics} \textbf{2005}, \emph{1}, 177\relax
\mciteBstWouldAddEndPuncttrue
\mciteSetBstMidEndSepPunct{\mcitedefaultmidpunct}
{\mcitedefaultendpunct}{\mcitedefaultseppunct}\relax
\EndOfBibitem
\bibitem{Laird-2010}
E.~A. Laird, J.~M. Taylor, D.~P. DiVincenzo, C.~M. Marcus, M.~P. Hanson, A.~C.
  Gossard, \emph{Physical Review B} \textbf{2010}, \emph{82}, 075403\relax
\mciteBstWouldAddEndPuncttrue
\mciteSetBstMidEndSepPunct{\mcitedefaultmidpunct}
{\mcitedefaultendpunct}{\mcitedefaultseppunct}\relax
\EndOfBibitem
\bibitem{Shi-2012}
Z.~Shi, C.~B. Simmons, J.~R. Prance, J.~K. Gamble, T.~S. Koh, Y.-P. Shim,
  X.~Hu, D.~E. Savage, M.~G. Lagally, M.~A. Eriksson, M.~Friesen, S.~N.
  Coppersmith, \emph{Physical Review Letters} \textbf{2012}, \emph{108},
  140503\relax
\mciteBstWouldAddEndPuncttrue
\mciteSetBstMidEndSepPunct{\mcitedefaultmidpunct}
{\mcitedefaultendpunct}{\mcitedefaultseppunct}\relax
\EndOfBibitem
\bibitem{Koh-2012}
T.~S. Koh, J.~K. Gamble, M.~Friesen, M.~A. Eriksson, S.~N. Coppersmith,
  \emph{Physical Review Letters} \textbf{2012}, \emph{109}, 250503\relax
\mciteBstWouldAddEndPuncttrue
\mciteSetBstMidEndSepPunct{\mcitedefaultmidpunct}
{\mcitedefaultendpunct}{\mcitedefaultseppunct}\relax
\EndOfBibitem
\bibitem{Kim-2014}
D.~Kim, Z.~Shi, C.~B. Simmons, D.~R. Ward, J.~R. Prance, T.~S. Koh, J.~K.
  Gamble, D.~E. Savage, M.~G. Lagally, M.~Friesen, S.~N. Coppersmith, M.~A.
  Eriksson, \emph{Nature} \textbf{2014}, \emph{511}, 70--74\relax
\mciteBstWouldAddEndPuncttrue
\mciteSetBstMidEndSepPunct{\mcitedefaultmidpunct}
{\mcitedefaultendpunct}{\mcitedefaultseppunct}\relax
\EndOfBibitem
\bibitem{Kim-2015}
D.~Kim, D.~R. Ward, C.~B. Simmons, D.~E. Savage, M.~G. Lagally, M.~Friesen,
  S.~N. Coppersmith, M.~A. Eriksson, \emph{Npj Quantum Information}
  \textbf{2015}, \emph{1}, 15004\relax
\mciteBstWouldAddEndPuncttrue
\mciteSetBstMidEndSepPunct{\mcitedefaultmidpunct}
{\mcitedefaultendpunct}{\mcitedefaultseppunct}\relax
\EndOfBibitem
\bibitem{Thorgrimsson-2017}
B.~Thorgrimsson, D.~Kim, Y.-C. Yang, L.~W. Smith, C.~B. Simmons, D.~R. Ward,
  R.~H. Foote, J.~Corrigan, D.~E. Savage, M.~G. Lagally, M.~Friesen, S.~N.
  Coppersmith, M.~A. Eriksson, \emph{Npj Quantum Information} \textbf{2017},
  \emph{3}, 32\relax
\mciteBstWouldAddEndPuncttrue
\mciteSetBstMidEndSepPunct{\mcitedefaultmidpunct}
{\mcitedefaultendpunct}{\mcitedefaultseppunct}\relax
\EndOfBibitem
\bibitem{Ferraro-2014}
E.~Ferraro, M.~De~Michielis, G.~Mazzeo, M.~Fanciulli, E.~Prati, \emph{Quantum
  Information Processing} \textbf{2014}, \emph{13}, 1155--1173\relax
\mciteBstWouldAddEndPuncttrue
\mciteSetBstMidEndSepPunct{\mcitedefaultmidpunct}
{\mcitedefaultendpunct}{\mcitedefaultseppunct}\relax
\EndOfBibitem
\bibitem{Ferraro-2015-qip}
E.~Ferraro, M.~De~Michielis, M.~Fanciulli, E.~Prati, \emph{Quantum Information
  Processing} \textbf{2015}, \emph{14}, 47--65\relax
\mciteBstWouldAddEndPuncttrue
\mciteSetBstMidEndSepPunct{\mcitedefaultmidpunct}
{\mcitedefaultendpunct}{\mcitedefaultseppunct}\relax
\EndOfBibitem
\bibitem{Ferraro-2017}
E.~Ferraro, M.~Fanciulli, M.~{De Michielis}, \emph{Quantum Information
  Processing} \textbf{2017}, \emph{16}, 277\relax
\mciteBstWouldAddEndPuncttrue
\mciteSetBstMidEndSepPunct{\mcitedefaultmidpunct}
{\mcitedefaultendpunct}{\mcitedefaultseppunct}\relax
\EndOfBibitem
\bibitem{DeMichielis-2015}
M.~De~Michielis, E.~Ferraro, M.~Fanciulli, E.~Prati, \emph{Journal of Physics
  A: Mathematical and Theoretical} \textbf{2015}, \emph{48}, 065304\relax
\mciteBstWouldAddEndPuncttrue
\mciteSetBstMidEndSepPunct{\mcitedefaultmidpunct}
{\mcitedefaultendpunct}{\mcitedefaultseppunct}\relax
\EndOfBibitem
\bibitem{Ferraro-2015-prb}
E.~Ferraro, M.~De~Michielis, M.~Fanciulli, E.~Prati, \emph{Physical Review B}
  \textbf{2015}, \emph{91}, 075435\relax
\mciteBstWouldAddEndPuncttrue
\mciteSetBstMidEndSepPunct{\mcitedefaultmidpunct}
{\mcitedefaultendpunct}{\mcitedefaultseppunct}\relax
\EndOfBibitem
\bibitem{Rotta-2016}
D.~Rotta, M.~De~Michielis, E.~Ferraro, M.~Fanciulli, E.~Prati, \emph{Quantum
  Information Processing Topical Collection} \textbf{2016}, \emph{15},
  2253--2274\relax
\mciteBstWouldAddEndPuncttrue
\mciteSetBstMidEndSepPunct{\mcitedefaultmidpunct}
{\mcitedefaultendpunct}{\mcitedefaultseppunct}\relax
\EndOfBibitem
\bibitem{Prati-npj}
D.~Rotta, F.~Sebastiano, E.~Charbon, E.~Prati, \emph{npj Quantum Information}
  \textbf{2017}, \emph{3}, 26\relax
\mciteBstWouldAddEndPuncttrue
\mciteSetBstMidEndSepPunct{\mcitedefaultmidpunct}
{\mcitedefaultendpunct}{\mcitedefaultseppunct}\relax
\EndOfBibitem
\bibitem{Wang-2017}
B.-C. Wang, G.~Cao, H.-O. Li, M.~Xiao, G.-C. Guo, X.~Hu, H.-W. Jiang, G.-P.
  Guo, \emph{Physical Review Applied} \textbf{2017}, \emph{8}, 064035\relax
\mciteBstWouldAddEndPuncttrue
\mciteSetBstMidEndSepPunct{\mcitedefaultmidpunct}
{\mcitedefaultendpunct}{\mcitedefaultseppunct}\relax
\EndOfBibitem
\bibitem{Ferraro-2018}
E.~Ferraro, M.~Fanciulli, M.~De~Michielis, \emph{Quantum Information
  Processing} \textbf{2018}, \emph{17}, 130\relax
\mciteBstWouldAddEndPuncttrue
\mciteSetBstMidEndSepPunct{\mcitedefaultmidpunct}
{\mcitedefaultendpunct}{\mcitedefaultseppunct}\relax
\EndOfBibitem
\bibitem{DasSarma-2016}
S.~Das~Sarma, R.~E. Throckmorton, Y.-L. Wu, \emph{Physical Review B}
  \textbf{2016}, \emph{94}, 045435\relax
\mciteBstWouldAddEndPuncttrue
\mciteSetBstMidEndSepPunct{\mcitedefaultmidpunct}
{\mcitedefaultendpunct}{\mcitedefaultseppunct}\relax
\EndOfBibitem
\bibitem{DasSarma-2017}
R.~E. Throckmorton, E.~Barnes, S.~Das~Sarma, \emph{Physical Review B}
  \textbf{2017}, \emph{95}, 085405\relax
\mciteBstWouldAddEndPuncttrue
\mciteSetBstMidEndSepPunct{\mcitedefaultmidpunct}
{\mcitedefaultendpunct}{\mcitedefaultseppunct}\relax
\EndOfBibitem
\bibitem{Wu-2017}
Y.-L. Wu, S.~{Das Sarma}, \emph{Physical Review B} \textbf{2017}, \emph{96},
  165301\relax
\mciteBstWouldAddEndPuncttrue
\mciteSetBstMidEndSepPunct{\mcitedefaultmidpunct}
{\mcitedefaultendpunct}{\mcitedefaultseppunct}\relax
\EndOfBibitem
\bibitem{XianWu-2014}
X.~Wu, D.~R. Ward, J.~R. Prance, D.~Kim, J.~K. Gamble, R.~T. Mohr, Z.~Shi,
  D.~E. Savage, M.~G. Lagally, M.~Friesen, S.~N. Coppersmith, M.~A. Eriksson,
  \emph{PNAS} \textbf{2014}, \emph{111}, 11938–11942\relax
\mciteBstWouldAddEndPuncttrue
\mciteSetBstMidEndSepPunct{\mcitedefaultmidpunct}
{\mcitedefaultendpunct}{\mcitedefaultseppunct}\relax
\EndOfBibitem
\end{mcitethebibliography}

\end{document}